%% file: FilterDesignDN_Arxiv.tex
\documentclass{article}
\usepackage{SKConferencePaperV0002}

\input{FilterDesignDN_TitleData}

\begin{document}

\maketitle

\input{FilterDesignDN_AbstractKeywords}

\section{Introduction}
\label{Introduction}
\input{FilterDesignDN_Introduction}

\section{Directed Networks:~Spectral Statistics}
\label{SpectralStatistics}
\input{FilterDesignDN_SpectralStatistics}

\section{Directed Networks:~~Filter Design}
\label{FilterDesignCriteria}
\input{FilterDesignDN_SimulationFigures}
\input{FilterDesignDN_DesignCriteria}

\section{Directed Networks:~~Simulations}
\label{NumericalSimulations}
\input{FilterDesignDN_NumericalSimulations}

\vfill
\section{Conclusion}
\vfill
\label{Conclusion}
\input{FilterDesignDN_Conclusion}

\clearpage
\begingroup
\bibliographystyle{IEEEtran}
\begin{spacing}{.92}
\bibliography{FilterDesignDN_Bibliography}
\end{spacing}
\endgroup

\end{document}

%% file: FilterDesignDN_TitleData.tex
\title{
   Optimal Filter Design for Consensus on Random Directed Graphs}

\name{\parbox{\linewidth}{\centering
Stephen Kruzick and Jos\'{e} M. F. Moura\\~\\
\textnormal{Carnegie Mellon University, Department of Electrical Engineering \\
5000 Forbes Avenue, Pittsburgh, Pennsylvania 15213}\vspace{-20pt}}
\thanks{
Stephen~Kruzick (skruzick@andrew.cmu.edu) and Jos\'{e}~M.~F.~Moura (moura@andrew.cmu.edu) are with the Department of Electrical and Computer Engineering at Carnegie Mellon University in Pittsburgh, PA, USA.  This work was supported by NSF grant \#~CCF~1513936. \vspace{-10pt}
}}
\address{}


%% file: FilterDesignDN_AbstractKeywords.tex
\begin{spacing}{.95}
\begin{abstract}
\vspace{1pt}
Optimal design of consensus acceleration graph filters relates closely to the eigenvalues of the consensus iteration matrix.  This task is complicated by random networks with uncertain iteration matrix eigenvalues.  Filter design methods based on the spectral asymptotics of consensus iteration matrices for large-scale, random \textit{undirected} networks have been previously developed both for constant and for time-varying network topologies.  This work builds upon these results by extending analysis to large-scale, constant, random \textit{directed} networks.  The proposed approach uses theorems by Girko that analytically produce deterministic approximations of the empirical spectral distribution for suitable non-Hermitian random matrices.  The approximate empirical spectral distribution defines filtering regions in the proposed filter optimization problem, which must be modified to accommodate complex-valued eigenvalues.  Presented numerical simulations demonstrate good results.  Additionally, limitations of the proposed method are discussed.\vspace{-5pt}

\end{abstract}

\begin{keywords}
graph signal processing, filter design, distributed average consensus, random graph, random matrix, spectral statistics, stochastic canonical equations
\end{keywords}
\end{spacing}

%% file: FilterDesignDN_Introduction.tex
\setlength{\abovedisplayskip}{2.5pt}
\setlength{\belowdisplayskip}{2.5pt}
\setlength{\abovedisplayshortskip}{1pt}
\setlength{\belowdisplayshortskip}{1pt}

\noindent Distributed average consensus, an iterative network process in which nodes compute the average of data spread among the network nodes through local communications, represents an important network agreement problem~\cite{ROlfSab2}.  This task appears in network-related applications such as processor load balancing~\cite{GCyb1}, sensor data fusion~\cite{LXia2}, distributed inference~\cite{SKar2}, and formation control or flocking of autonomous agents~\cite{ROlfSab1}.  Every network node begins with an initial scalar data element, collected into a vector $\mathbf{x}_0$ in which each row corresponds to a node.  They each maintain a scalar state variable over time, collected into a vector $\mathbf{x}_n$ at time iteration $n$.  At each iteration, the nodes communicate with neighboring nodes in the network and update their state according to a linear combination of locally available data.  This implements the dynamics 
\begin{equation}\label{ConsensusSystem}
\mathbf{x}_n=W\left(\mathcal{G}\right)\mathbf{x}_{n-1}
\end{equation}
where $W\left(\mathcal{G}\right)$ is the consensus iteration weight matrix, which
 must respect the network graph structure $\mathcal{G}$~\cite{SKar2}.  For a given connected network topology, there are many possible iteration weight matrices.  Consensus will be reached provided the following conditions hold,
\begingroup
\thinmuskip=.1\thinmuskip
\medmuskip=.1\medmuskip
\thickmuskip=.1\thickmuskip
\begin{equation}\label{ConsensusCond}
{\boldsymbol \ell}^\top W={\boldsymbol \ell}^\top, \enskip W\mathbf{1}=\mathbf{1}, \enskip \rho\left(W-J_{\boldsymbol \ell}\right)<1, \enskip J_{\boldsymbol \ell}=\mathbf{1}{\boldsymbol \ell}^\top/{\boldsymbol \ell}^\top\mathbf{1}
\end{equation}
\endgroup
where $\rho$ is the spectral radius and $J_{\boldsymbol \ell}$ is the matrix that produces consensus to an average weighted by ${\boldsymbol \ell}$, the left eigenvector of $W$ corresponding to eigenvalue $\lambda=1$~\cite{SKar2}.  \mbox{That is,}
\begin{equation}
\lim_{n\rightarrow\infty} \mathbf{x}_n=J_{\boldsymbol \ell} \mathbf{x}_0.
\end{equation}
at an exponential rate governed by $\ln\left(\rho\left(W-J_{\boldsymbol \ell}\right)\right)$~\cite{SKar2}.

Design of consensus dynamics for fast convergence can be approached in several ways, including design of the weight matrix $W$ given the network topology $\mathcal{G}$~\cite{LXia1} as well as design of the network topology $\mathcal{G}$ under constraints given a weight matrix scheme $W\left(\mathcal{G}\right)$~\cite{SKar1}.  A third approach involves designing filters that are applied to the system state at each node.  Some example filter design methods proposed for various scenarios can be found in~\cite{SSun1,EKok1,ASan4,EMon1,ALou1,SApe1,FGam1}.  For this paper, each node implements~\eqref{ConsensusSystem} at each iteration.  Additionally, a degree $d$ filter will periodically be applied to the state value recorded at each node every $d$ iterations.  For filter coefficients $\left\{a_n\right\}_{n=0}^{n=d}$, this may be expressed as
\begingroup
\thinmuskip=.1\thinmuskip
\medmuskip=.1\medmuskip
\begin{equation}
\mathbf{x}_n :={\textstyle\sum_{ k=0}^{ k=d}} a_k\mathbf{x}_{n-d+k}, \quad n\equiv 0~ \left(\textrm{mod}~d\right).
\end{equation}
\endgroup
Furthermore, for constant network topology $\mathcal{G}$, this can be expressed as $\mathbf{x}_n=p\left(W\left(\mathcal{G}\right)\right)\mathbf{x}_{n-d}$ where $p$ is a polynomial graph filter with coefficients $\left\{a_n\right\}_{n=0}^{n=d}$.  The filter should be designed to optimize the convergence rate by minimizing the convergence factor $\frac{1}{d}\ln\left(p\left(W\right)-J_{\boldsymbol\ell}\right)$.  For constant, random networks,~\cite{SKru1,SKru2,SKru3,SKru4} find deterministic approximations for the empirical spectral distribution of consensus iteration matrices for large-scale random, symmetric networks and use this for consensus acceleration filter design, while~\cite{SKru5} handles networks with time-varying (switching) network topology.

This paper extends previous work connecting spectral asymptotics to graph filter design for consensus acceleration by examining the filter design problem in the context of constant (not time-varying), random networks of large-scale.  Section~\ref{SpectralStatistics} discusses a random matrix theory method useful for analyzing the asymptotics of the empirical spectral distribution of non-Hermitian random matrices, demonstrating its application to a consensus iteration matrix for an example directed random graph model.  Section~\ref{FilterDesignCriteria} poses a filter design optimization problem for directed random networks that selects eigenvalue filtering regions based on the deterministic approximations to the empirical spectral distributions of the iteration matrix, as derived in the preceding section.  Section~\ref{NumericalSimulations} supports the proposed design method with numerical simulation results comparing filtered convergence rates.  Finally, Section~\ref{Conclusion} provides concluding analysis.

%% file: FilterDesignDN_SpectralStatistics.tex
\setlength{\abovedisplayskip}{2pt}
\setlength{\belowdisplayskip}{2pt}
\setlength{\abovedisplayshortskip}{1pt}
\setlength{\belowdisplayshortskip}{1pt}

\newcommand{\scval}{.8}

\noindent Consider a potentially non-Hermitian $N\times N$ random matrix $\Xi_N$ with potentially complex eigenvalues $\lambda_i\left(\Xi_N\right)$ for $i=1,\ldots, N$.  The empirical spectral distribution and corresponding empirical spectral density, functions of the real and imaginary components of a complex parameter, encapsulate the eigenvalue information.  These functions are, respectively, given by
\begingroup
\thinmuskip=.1\thinmuskip
\medmuskip=.1\medmuskip
\thickmuskip=.1\thickmuskip
\begin{align}
\kern-.5em
F_{\Xi_N}\left(x,y\right)&=\scalebox{\scval}{$\displaystyle\frac{1}{N}\sum_{i=1}^{i=N}$}\chi\left(x\leq\Re{\lambda_i\left(\Xi_N\right)},y\leq\Im{\lambda_i\left(\Xi_N\right)}\right) \kern-.1em\\
\kern-.5em f_{\Xi_N}\left(x,y\right)&=\scalebox{\scval}{$\displaystyle\frac{1}{N}\sum_{i=1}^{i=N}$}\delta(x-\Re{\lambda_i\left(\Xi_N\right)},y-\Im{\lambda_i\left(\Xi_N\right)})\kern-.1em
\end{align}
\endgroup
where $\chi$ is an indicator function and $\delta$ is the Dirac delta function. Although these are random functions, the limiting behavior of the empirical spectral distribution sometimes admits a deterministic characterization that  provides useful information through theorems from random matrix theory.  Classic examples include the Wigner semicircular law~\cite{EWig1}, Marchenko-Pastur law~\cite{VMar1,RCou1}, and Girko circular~law~\cite{VGir1}.

Matrices arising from random networks necessitate analysis methods that handle random matrix models with structure that require certain entries to be zero.  While many random matrix theory tools focus on matrices with independent, identically distributed entries, theorems called stochastic canonical equations described by Girko~\cite{VGir1} accommodate non-identically distributed entries and, thus, zeros forced by graph structure.  These methods involve solving a system of equations that depends on the random matrix model to find a deterministic equivalent for the empirical spectral distribution of a large-scale  matrix. For symmetric matrices, Girko's K1 Equation was applied to network adjacency matrices and consensus iteration matrices in~\cite{KAvr1,SKru1,SKru2}, information that was then used to inform filter design optimization problems for consensus acceleration in~\cite{SKru3,SKru4,SKru5}.  For the non-symmetric random network models on which this paper focuses, a much more complex method shown, in abridged form, as Theorem~\ref{GirkoK25} (Girko's K25 Equation) is required to perform analysis.  

\vspace{-10pt}
\begin{theorem}[Girko's K25 Equation (abr.)~\cite{VGir1}]\label{GirkoK25}
Let $\Xi_N$ be a family of complex-valued $N\times N$ random matrices with independent entries that satisfy several regularity conditions. (See Theorem 25.1 of~\cite{VGir1} for the full list.)  Let $\Xi_N$ have expectation $B_N=\Exp{\Xi_N}$ and centralization $H_N=\Xi_N-B_N$ with entry variance $\sigma_{N,ij}^2=\operatorname{E}[|\left(H_N\right)_{ij}|^2]$.  Then 
\begin{equation}
\lim_{\beta\rightarrow 0^+} { \lim_{N\rightarrow \infty} {\left\|F_{\Xi_N}\left(x,y\right)-\widehat{F}_{\Xi_N,\beta}\left(x,y\right)\right\|}}=0
\end{equation}
almost surely, where

\noindent\parbox{\linewidth}{
\begingroup
\thinmuskip=.1\thinmuskip
\medmuskip=.1\medmuskip
\begin{equation}
\kern-.6em\scalebox{.9}{$\displaystyle\frac{\partial^2\widehat{F}_{\Xi_N,\beta}\left(t,s\right)}{\partial x \partial y}
\kern-.45em =\kern-.45em\left\{\kern-.8em\begin{array}{cc} -\kern-.1em\frac{1}{4\pi}\kern-.1em\int_\beta^\infty \kern-.3em \left(\kern-.3em\frac{\partial^2}{\partial t^2}\kern-.1em+\kern-.1em\frac{\partial^2}{\partial s^2}\kern-.3em\right)\kern-.1em m_N\left(u,t,s\right)du & \kern-.5em (t,s)\kern-.3em\notin\kern-.3em G \\ \!\!\!\!\! 0 & \kern-.5em \left(t,s\right) \kern-.3em\in\kern-.3em G \end{array}\right.\kern-.8em$} \label{GirkoK25_Dens}
\end{equation}
\endgroup}
(with the region $G$ defined below) and
\begingroup
\thinmuskip=.05\thinmuskip
\medmuskip=.05\medmuskip
\begin{equation}
\begin{aligned}
m&_N\left(u,t,s\right)\kern-.2em = \kern-.2em \frac{1}{N}\tr\left[\left(C_1\left(u,s,t\right)+\ldots \vphantom{\left(B-(t+\im s)I_N\right)C_2\left(s,t\right)^{-1}\left(B_N-(t+\im s)I\right)^{*}}\right.\right.  \\
&\kern-.2em \left.\vphantom{\frac{1}{N}}\left. \vphantom{(C_1\left(s,t\right)+}\left(B_N-(t+\im s)I\right)C_2\left(u,s,t\right)^{-1}\left(B_N-(t+\im s)I\right)^{*}\right)^{\kern-.2em -1}\right]
\end{aligned}
\end{equation}
\endgroup
for $u>0$.  The matrices $C_1\left(u,s,t\right)$ and $C_2\left(u,s,t\right)$ are diagonal matrices with entries that satisfy the system of equations
\begingroup
\thinmuskip=.1\thinmuskip
\medmuskip=.1\medmuskip
\begin{align}
&\begin{aligned}\kern-1em
(C_1&)_{kk}\left(u,s,t\right)=u+\scalebox{\scval}{$\displaystyle\sum_{j=1}^{j=N}$}\sigma_{N,kj}^2\left[\left(C_2\left(u,s,t\right)+\ldots\vphantom{\left(B_N-(t+s\im)I\right)^*C_1\left(u,s,t\right)^{-1}\left(B_N-(t+s\im)I\right)}\right.^{\vphantom{-1}}\right. \\
&\left.\left.\vphantom{C_2\left(y,s,t\right)+}\left(B_N-(t+s\im)I\right)^*C_1\left(u,s,t\right)^{-1}\left(B_N-(t+s\im)I\right)\right)^{\kern-.2em -1}\right]_{\kern-.2em jj \kern-2em}
\end{aligned}\label{GirkoK25_Sys1}
\\
&\begin{aligned}\kern-1em
(C_2&)_{\ell\ell}\left(u,s,t\right)=1+\scalebox{\scval}{$\displaystyle\sum_{j=1}^{j=N}$}\sigma_{N,j\ell}^2\left[\left(C_1\left(u,s,t\right)+\ldots\vphantom{\left(B_N-(t+s\im)I\right)C_2\left(y,s,t\right)^{-1}\left(B_N-(t+s\im)I\right)^*} \right.^{\vphantom{-1}}\right. \\
&\left.\left.\vphantom{C_1\left(y,s,t\right)+}\left(B_N-(t+s\im)I\right)C_2\left(u,s,t\right)^{-1}\left(B_N-(t+s\im)I\right)^*\right)^{\kern-.2em -1}\right]_{\kern-.2em jj \kern-2em}
\end{aligned}\label{GirkoK25_Sys2}
\end{align}
\endgroup
for $k,\ell=1,\ldots N$.  There exists a unique solution to this system of equations among real positive analytic functions in $u>0$.  The region $G$ is given by
\begingroup
\thinmuskip=.1\thinmuskip
\medmuskip=.1\medmuskip
\begin{equation}
G=\left\{ (t,s) \middle| \limsup_{\beta\rightarrow 0^+}\limsup_{N\rightarrow\infty}\left|\frac{\partial}{\partial \beta}m_N\left(\beta,t,s\right)\right|<\infty\right\}.
\end{equation}
\endgroup
\end{theorem}

Thus, for large-scale random matrices that satisfy the conditions of Theorem~\ref{GirkoK25}, an approximation to the empirical spectral density can be achieved in which the pointwise error converges to zero almost surely.  The methods presented in Section~\ref{FilterDesignCriteria} employ this approximate distribution and the corresponding density to define regions for filter response optimization.  Towards that goal, the application of Girko's method to consensus iteration matrices that arise from directed random networks must first be justified, described, \mbox{and demonstrated.}

In order to obtain a deterministic approximation of the empirical spectral distribution, the system of equations \mbox{\eqref{GirkoK25_Sys1}-\eqref{GirkoK25_Sys2}} must be solved numerically for numerous $u$ values ranging from a small value of $\beta>0$ to a very large value such that the integral in \eqref{GirkoK25_Dens} can be approximated.  This must be accomplished for all $t+s\im$ for which the value of $F_{\Xi_N,\beta}(t,s)$ is required.  Hence, the system must be solved for numerous points in a three dimensional region.  This represents a significant computational burden that must be accomplished offline in advance of network deployment using foreknowledge of the network iteration matrix distribution.  Note that the system can be solved via an iterative fixed point search, similar to the procedure in~\cite{SKru1,SKru2}, because a unique solution exists.  Also note that the system has $2N$ equations, where $N$ is typically large, presenting a computational challenge.  When possible, reduction of the system of equations~\eqref{GirkoK25_Sys1}-\eqref{GirkoK25_Sys2} via diagonalization addresses this problem, as done in~\cite{SKru1,SKru2} for symmetric matrices and Girko's K1 equation.  The following example analyzes the application of this equation for deterministically approximating the empirical spectral distribution of a consensus iteration matrix for a non-symmetric stochastic block model (briefly, proofs and derivations omitted).  These results, and the design methods produced in Section~\ref{FilterDesignCriteria} were used to produce the simulation results shown in Section~\ref{NumericalSimulations}.

\vspace{-10pt}
\begin{example}[Stochastic Block Model]  Consider a directed stochastic block model network~\cite{KAvr1} with $M$ populations of equal size $S$ such that there are $N=MS$ nodes.  Let each pair of distinct nodes in populations $1\leq i,j \leq M$ connect with probability $\Theta_{ij}=\Theta_{ji}$.  Note that the connection probability is symmetric, but the outcomes for each link direction are independent.  Furthermore, for each pair of populations $1\leq i,j \leq M$, let there be some automorphism on the populations that preserves $\Theta$ and maps population $i$ to population $j$.  This produces node transitivity on the distribution (but not outcome) of the random graph.

For filter design in this paper, an estimate of the spectral distribution for the consensus iteration matrix (based on the directed, row normalized Laplacian through $W_{\!N}\!\!\!=\!\!\!I\!\!\!-\!\!\!\alpha \widehat{L}_R$\mbox{) is} required, which can be accomplished through a scaled adjacency matrix $\Xi_N=1/\gamma A_N\left(\mathcal{G}\right)$ with $\gamma=\rho\left(\operatorname{E}\left[A_N\right]\right)$ under certain conditions. The approximate distribution for $W_N$ will be derived from that found for $\Xi_N$ through
\begin{equation}
\widehat{f}_{W_N,\beta}\left(x,y\right)=\tfrac{1}{\alpha^2}\widehat{f}_{\Xi_N,\beta}\left(\tfrac{(x-1)}{\alpha}+1,\tfrac{y}{\alpha}\right).
\end{equation}
The approach for solving Girko's equation for this model is briefly described, with full details omitted for space.  For any random iteration matrix model arising node-transitive random network, the diagonal matrices $C_1,C_2$ from Theorem~\ref{GirkoK25} must become scalar matrices $C_1=c_1I, C_2=c_2I$ and the variance row sums and column sums must be equal.  This allows simplified computation via the trace by summing both sides of~\eqref{GirkoK25_Sys1}-\eqref{GirkoK25_Sys2}.  In the right side expressions, a trace emerges that then can be written in terms of the mean matrix eigenvalues.  Further reduction occurs because $B_N$ is real and symmetric for this case. The resulting equations appear below.
\begingroup
\thinmuskip=.1\thinmuskip
\medmuskip=.1\medmuskip
\begin{align}
&\begin{aligned}\kern-.4em
c_1\kern-.2em=&\tfrac{1}{N}\tr\left(C_1\right)=u+\left(\scalebox{\scval}{$\displaystyle\frac{1}{N}{\sum_{k=1}^{k=N}}$}\sigma_{N,kj}^2\right)\times \ldots \\
&\scalebox{\scval}{$\displaystyle\sum_{i=1}^{i=N}$}\left({c_2+1/c_1 \left(\lambda_i\left(B_N\right)^2-2t\lambda_i\left(B_N\right)+\left|t+\im s\right|^2\right)}\right)^{\kern-.3em -1 \kern-.4em}
\end{aligned}\\
&\begin{aligned}\kern-.4em
c_2\kern-.2em=&\tfrac{1}{N}\tr\left(C_2\right)=1+\left(\scalebox{\scval}{$\displaystyle\frac{1}{N}{\sum_{\ell=1}^{\ell=N}}$}\sigma_{N,j\ell}^2\right)\times \ldots \\
&\scalebox{\scval}{$\displaystyle\sum_{i=1}^{i=N}$}\left({c_1+1/c_2 \left(\lambda_i\left(B_N\right)^2-2t\lambda_i\left(B_N\right)+\left|t+\im s\right|^2\right)}\right)^{\kern-.3em -1 \kern-.4em}
\end{aligned}
\end{align}  
\endgroup
These equations can then be approximately solved at every necessary value of $(u,t,s)$ via an iterative fixed point search as done for Girko's K1 equation in~\cite{SKru1,SKru2}.  Numerical integration is then conducted to find the density.
\end{example}

%% file: FilterDesignDN_SimulationFigures.tex
\begin{figure*}

\begin{floatrow}[3]

\ffigbox[.30\textwidth]
{\centering
\includegraphics[width=.925\linewidth]{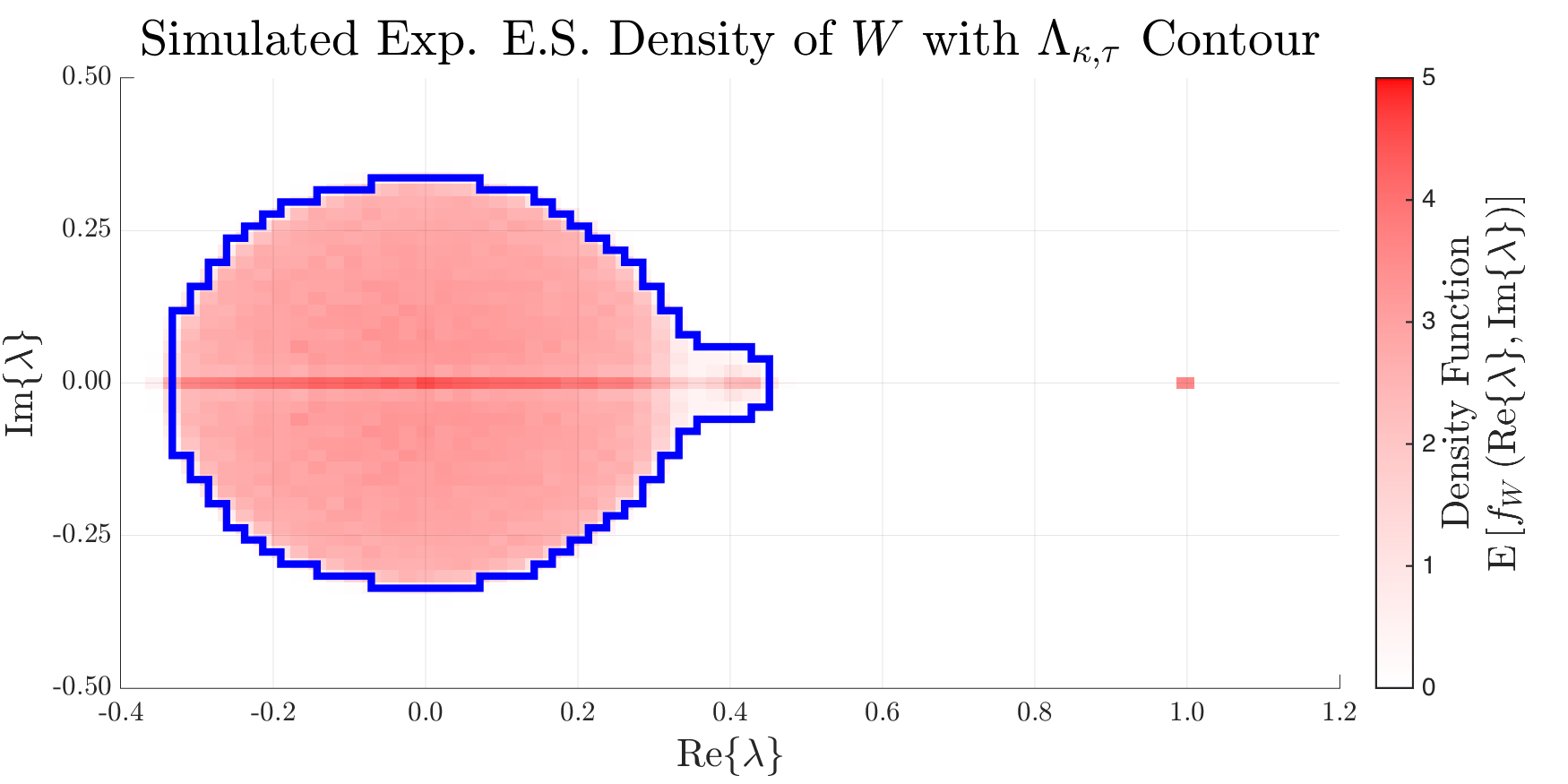}}
{\caption{Exp. empirical spectral density $\operatorname{E}\left[f_{W_N}\right]$ of $W_N$ ($1000$ Monte-Carlo trials) with outline of $\Lambda_{\kappa,\tau}$ (blue) dervied from $\widehat{f}_{W_N,\beta}$ (see Figure~\ref{NumSimB})
\vspace{-20pt}
}\label{NumSimA}}

\hspace{.01\textwidth}

\ffigbox[.30\textwidth]
{\includegraphics[width=.925\linewidth]{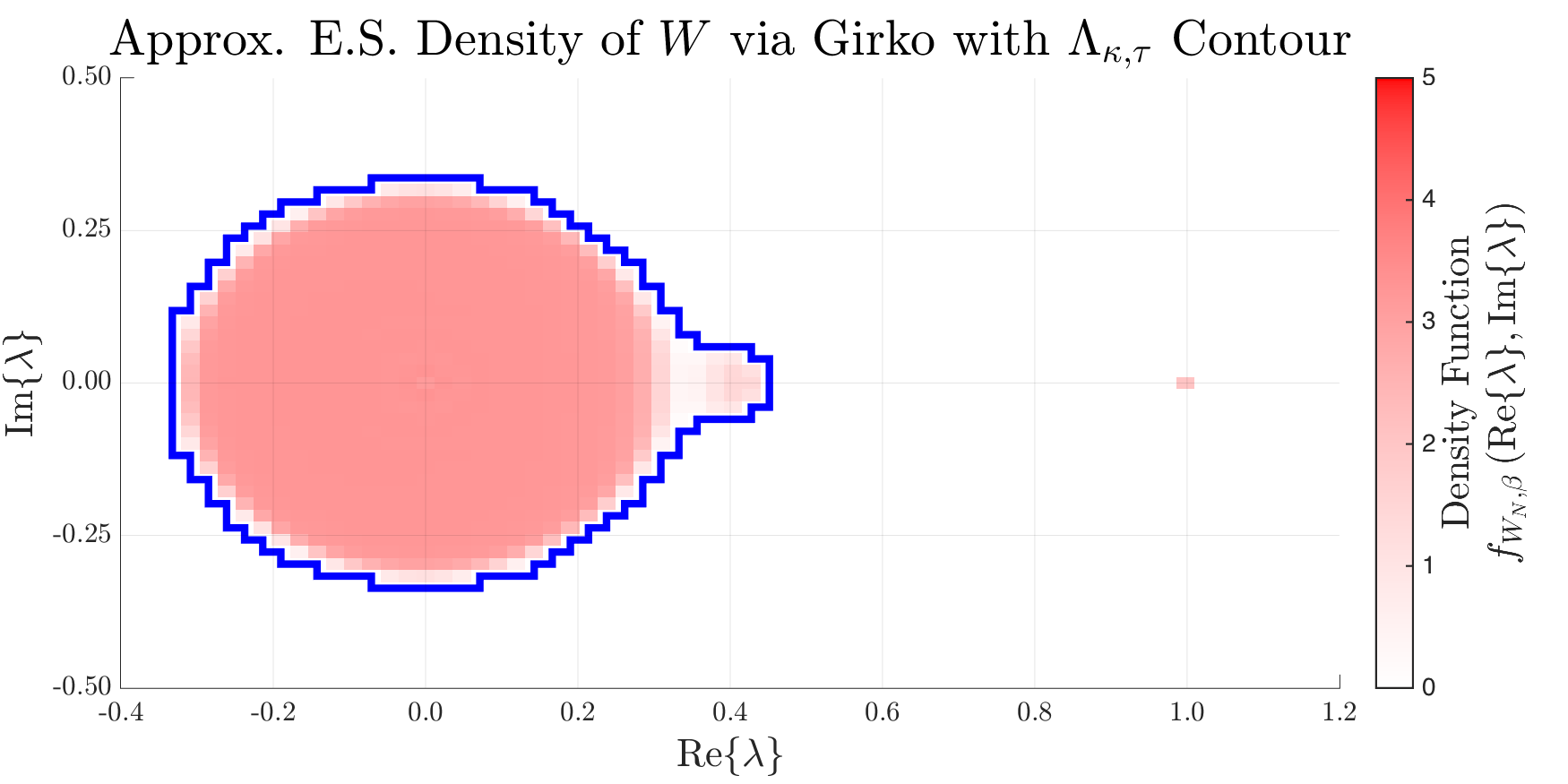}}
{\caption{
Deterministic approx. density $\widehat{f}_{W_N,\beta}$ computed from $\widehat{f}_{\Xi_N,\beta}$ via Girko's theorem as described in Section~\ref{SpectralStatistics} along with $\Lambda_{\kappa,\tau}$ outline (blue)
\vspace{-20pt}
}\label{NumSimB}}

\hspace{.01\textwidth}

\ffigbox[.30\textwidth]
{\includegraphics[width=.925\linewidth]{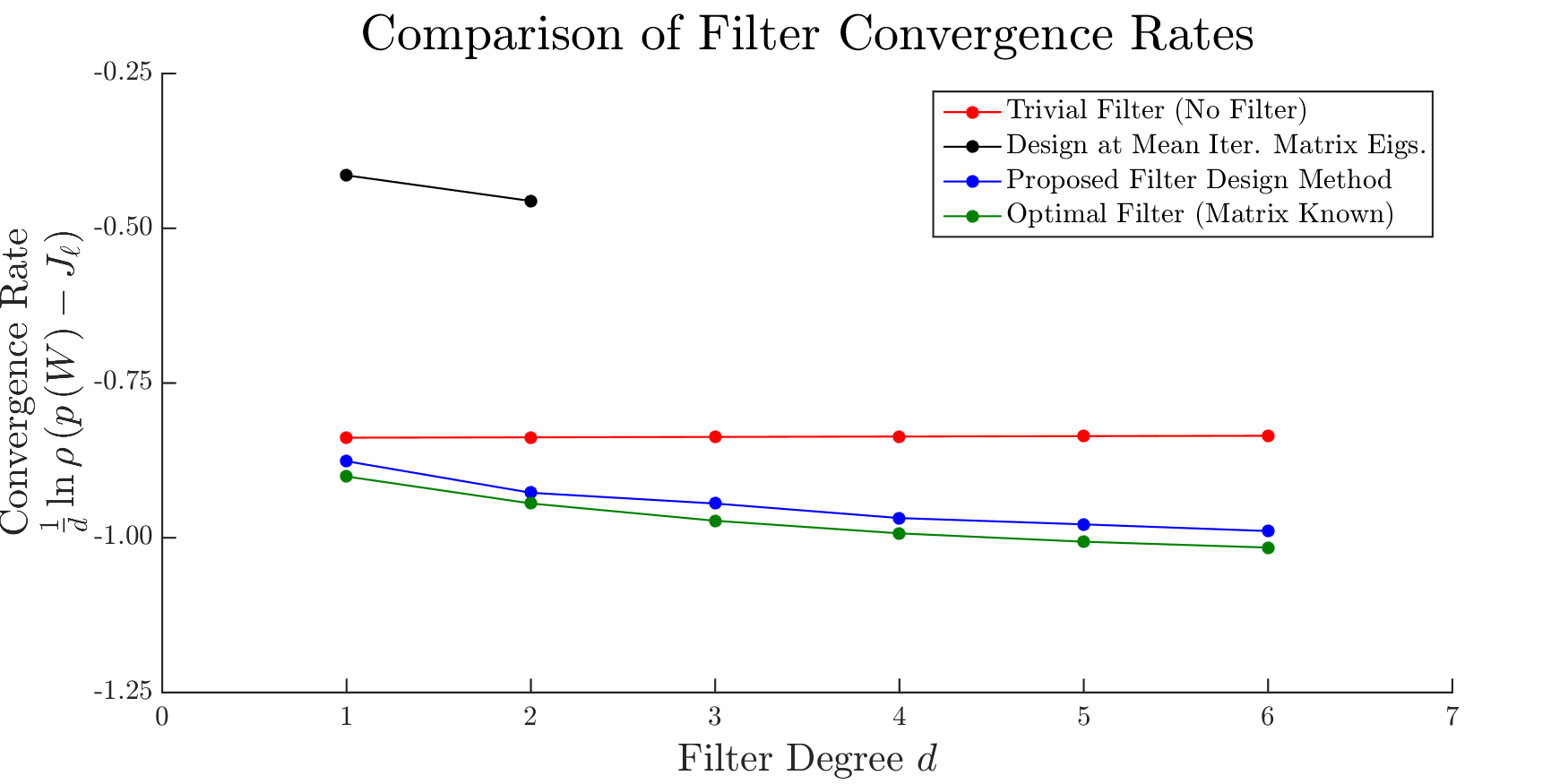}}
{\caption{Worst case filter response (log scale, per iteration) compared for several filters (proposed design in blue), determines the convergence rate
\vspace{-20pt}
}\label{NumSimC}}

\end{floatrow}

\end{figure*}

%% file: FilterDesignDN_DesignCriteria.tex
\noindent This section considers consensus acceleration filter design for a random network with constant (non-time-varying) topology described by a random graph $\mathcal{G}$ under a fixed scheme for determining the consensus iteration matrix from the random graph topology.  Thus, the consensus iteration matrix is drawn once from a random matrix distribution  and used for all time iterations.  This section presents filter design criteria for this scenario.  Assume that a deterministic equivalent for the empirical spectral distribution of the corresponding random iteration matrix is available, such as through the method in Section~\ref{SpectralStatistics}.  Recall that there are many potential choices for consensus iteration matrices that satisfy the consensus convergence conditions~\eqref{ConsensusCond}.
In particular, this paper uses iteration matrix scheme 
\begin{equation}
W\left(\mathcal{G}\right)=I-\alpha \widehat{L}_R\left(\mathcal{G}\right)
\end{equation}
based on the directed, row normalized Laplacian matrix
\begin{equation}
\widehat{L}_R\left(\mathcal{G}\right)=I-D\left(\mathcal{G}\right)^{-1}A\left(\mathcal{G}\right)
\end{equation}
where $A\left(\mathcal{G}\right)$ is the network graph adjacency matrix, $D\left(\mathcal{G}\right)$ is the diagonal matrix of node outdegrees, and $\alpha$ is chosen to satisfy the spectral radius condition in~\eqref{ConsensusCond}.  While this produces a weighted average (unlike unnormalized Laplacian based weights for symmetric graphs), this choice is easier to analyze through Girko's methods and can have convergence rate advantages~\cite{SKru3,SKru4}. Unlike the undirected context in which the weighting could be corrected through pre-multiplication if each node knows its degree \cite{SKru3}, this weighted average must be accepted as the left-eigenvector ${\boldsymbol \ell}$ is not easily computable from basic degree information.  

Consensus acceleration filters seek to minimize the following expression for the convergence rate, where $W$ is the random, constant (non-time-varying) iteration matrix.
\begin{equation}
\lim_{k\rightarrow \infty}\left\|p\left(W\right)^k-J_{\boldsymbol\ell}\right\|_2^{1/k}=\lim_{k\rightarrow\infty}\left\|\left(p\left(W\right)-J_{\boldsymbol\ell}\right)^k\right\|_2^{1/k}
\end{equation}
By Gelfand's formula~\cite{PLax1}, this reduces to the spectral radius $\rho\left(p\left(W\right)-J_{\boldsymbol\ell}\right)$.  Note that the eigenvalues of $p\left(W\right)$ are precisely $p\left(\lambda_i\left(W\right)\right)$ for each eigenvalue $\lambda_i\left(W\right)$ of $W$ by the spectral mapping theorem~\cite{PLax2}.  Let the eigenvalues of $W$ be $\lambda_1\left(W\right),\ldots, \lambda_{N}\left(W\right)$, where $\lambda_N\left(W\right)=1$, and be ordered such that $\left|\lambda_i\left(W\right)\right|\leq\left|\lambda_j\left(W\right)\right|$ for $1\leq i<j \leq N$.  Subtracting $J_{\boldsymbol \ell}$ removes the consensus eigenvalue $\lambda_N\left(W\right)=1$.  Thus $\rho\left(p\left(W\right)-J_{\boldsymbol\ell}\right)$ is the maximum absolute value of $p\left(\lambda_i\left(W\right)\right)$ for $1\leq i \leq N-1$.

Therefore, the worst case consensus convergence rate with periodic filtering can be approximately minimized through the following minimax optimization problem, where $P_d$ is the space of (real coefficient) polynomials of degree at most $d$.
\begin{equation}\label{Problem1}
\kern-.1em\begin{gathered}
\begin{aligned}
\min_{p\in P_d} { \max_{\lambda\in \Lambda_{\kappa, \tau}} {\left|p\left(\lambda\right)\right|}} \qquad
\st \enskip p\left(1\right)=1 
\end{aligned}\\
\Lambda_{\kappa,\tau}=\left\{\left|\lambda-1\right|>\kappa \middle| \widehat{f}_{W_N,\beta}\left(\Re{\lambda},\Im{\lambda}\right)>\tau\right\}
\end{gathered}
\end{equation}
The filtering region defined by the set $\Lambda_{\kappa,\tau}$ is determined by a deterministic approximation for the empirical spectral density, which can be obtained as described in Section~\ref{SpectralStatistics}.  This substitutes for knowledge of the true set of eigenvalues from the random iteration matrix.  Because computation of $\widehat{f}_{W_N,\beta}$ involves numerical computation of integrals and limits, it is necessary to threshold the result to define the filtering region.  The parameter $\tau$ (small value chosen) fills this role, while the parameter $\kappa$ (small value chosen) provides a transition region around the equality constraint.  Note that some computation could be saved by simply transforming the complement of the region $G$ from Theorem~\ref{GirkoK25}, but the above formulation is more directly analogous to that from~\cite{SKru3,SKru4}.

By introducing $\varepsilon$ to bound the maximum filter response magnitude squared and examining the response only at sample points $\Lambda_S\subseteq \Lambda_{\kappa,\tau}$, an approximate solution to~\eqref{Problem1} can be found by solving the following problem.
\begin{equation}\label{Problem2}
\kern-.4em\begin{gathered}
\begin{aligned}
\smash{\min_{p\in P_d,\varepsilon}} \enskip  \varepsilon \qquad
\st \enskip & p\left(1\right)=1 \\
& \left|p\left(\lambda_i\right)\right|^2<\varepsilon \quad \textrm{for all}~ \lambda_i\in \Lambda_S
\end{aligned}\\
\end{gathered}
\end{equation}
The precise scheme to determine $\Lambda_S$ is of little importance, but it should be sufficient to approximately capture the structure of $\Lambda_{\kappa,\tau}$ in discretized form.  Collecting the filter coefficients $\left\{a_n\right\}_{n=0}^{n=d}$ into a vector $\mathbf{a}$, the optimization problem~\eqref{Problem2} can be recast as
\begin{equation}\label{Problem3}
\kern-.4em\begin{aligned}
\smash{\min_{\mathbf{a}\in \mathbb{R}^{d+1}\!,\varepsilon}} \enskip \varepsilon \qquad
\mrlap{\st}\hphantom{\min_{\mathbf{a}}} \enskip & \mathbf{1}^\top\mathbf{a}=1 \\
&\mathbf{a}^\top Q\left(\lambda_{i}\right)\mathbf{a} < \varepsilon
\quad \textrm{for all}~ \lambda_i\in \Lambda_S
\end{aligned}
\end{equation} where $Q\left(\lambda_i\right)$ is the real, positive semidefinite matrix
\begin{equation}
Q\left(\lambda_i\right)=\scalebox{1}{$\displaystyle\frac{1}{2}$}\left(V\left(\lambda_i\right)^*V\left(\lambda_i\right)+V\left(\overline{\lambda}_i\right)^*V\left(\overline{\lambda}_i\right)\right)
\end{equation}
and $V(\lambda_i)$ is the Vandermonde row vector
\begin{equation}
V\left(\lambda_i\right)=\left[\lambda_i^0,\ldots,\lambda_i^d\right].
\end{equation}
This optimization problem has linear objective function with positive semidefinite quadratic contraints for each sample point (QCLP), and, thus, is convex~\cite{SBoy1}.  This approach mirrors that from~\cite{SKru3,SKru4} for symmetric matrices, where the real-valued eigenvalues produce linear inequality constraints.  Section~\ref{NumericalSimulations} shows numerical simulation results that demonstrate improved filters computed through this method.

%% file: FilterDesignDN_NumericalSimulations.tex
\noindent This section shows example results, displayed in Figures~\mbox{\ref{NumSimA}-\ref{NumSimB}}, for a directed stochastic block model with $N=600$ nodes divided into $M=6$ populations each with $S=100$ nodes.  For this simulation, the independent link probabilities $\Theta_{ij}$ between two disinct nodes in each ordered pair of populations $i,j$ are $\Theta_{ij}=0.05$ for $i=j$ and $\Theta_{ij}=0.01$ for $i\neq j$.
The simulation uses $W=I-\alpha \widehat{L}_R$ for the consensus iteration matrix ($\alpha=1$).

Figure~\ref{NumSimA} shows the expected empirical spectral distribution, averaged over 1000 Monte-Carlo trials (independently drawn random networks), in heat map form along with the outline of the region $\Lambda_{\kappa,\tau}$ ($\kappa=10^{-2}$, $\tau=10^{-4}$) isolated from the approximate density function $\widehat{f}_{W_N,\beta}$
($\beta=10^{-6}$).  The heatmap for the approximate density function $\widehat{f}_{W_N,\beta}$ appears in Figure~\ref{NumSimB}, also with the outline of $\Lambda_{\kappa,\tau}$.  

Figure~\ref{NumSimC} plots the expected worst case exponential convergence rate per iteration ${\frac{1}{d}}\ln\left(\rho\left(p\left(W\right)-J_{\boldsymbol\ell}\right)\right)$ of the filtered consensus system, averaged over 1000 Monte-Carlo trials (independently drawn random networks) for filter degrees $d=1,\ldots,6$. The proposed filter design method is compared against the trivial filter (no filtering), a filter designed to minimize response at the eigenvalues of the mean iteration matrix \cite{EKok1} (only for $d\leq K-1$ where $K$ is the number of distinct mean matrix eigenvalues, for this simulation $K=3$), and a filter designed with oracle knowledge of the true eigenvalues (optimal).  The proposed method (blue) performs nearly as well as the optimal filter (green), providing strong support.  Note that attempting to optimize at only the eivenvalues of the mean iteration matrix (black) performs poorly in this case because the eigenvalues spread over a wide region.

%% file: FilterDesignDN_Conclusion.tex
\noindent This paper considered graph filter design for accelerated consensus via periodic filtering on large-scale, directed random networks that have random non-symmetric consensus iteration matrices with tractable spectral asymptotics.  Similar to the approach previously taken for undirected random network models with symmetric random iteration matrices, this work first examined tools from random matrix theory to compute deterministic approximations for the empirical spectral distribution of the random consensus iteration matrix.  Subsequently, filter design criteria were proposed that employ this information to define filtering regions, resulting in an optimization problem to approximately minimize the convergence rate of the filtered consensus dynamics.  Numerical simulations demonstrated that filters designed via the proposed method achieve convergence rates close to the optimal convergence rate with full knowledge of the iteration matrix eigenvalues.  This approach has limitations, including complex numerical computations in Girko's K25 equation, restrictions on the random matrix models to which Girko's K25 equation can be efficiently applied, and a difficult to correct weighted average consensus.  Continuing work focuses on extending analysis to additional directed random network models and on handling time-varying random networks.